\documentclass[aps,twocolumn,showpacs,superscriptaddress]{revtex4}
\usepackage{epsfig}

\begin{document}


\title{Influence of magnetic field on paramagnetic-ferromagnetic transition
 in La$_{1-x}$Ca$_{x}$MnO$_{3}$ ($x\approx 0.25$) crystal:
 ultrasonic and transport studies}

\author{B. I. Belevtsev}
\email[]{belevtsev@ilt.kharkov.ua}
\affiliation{B. Verkin Institute for Low Temperature Physics and
Engineering, National Academy of Sciences, pr. Lenina 47, Kharkov
61103, Ukraine}

\author{G. A. Zvyagina}
\affiliation{B. Verkin Institute for Low Temperature Physics and
Engineering, National Academy of Sciences, pr. Lenina 47, Kharkov
61103, Ukraine}

\author{K. R. Zhekov}
\affiliation{B. Verkin Institute for Low Temperature Physics and
Engineering, National Academy of Sciences, pr. Lenina 47, Kharkov
61103, Ukraine}

\author{I. G. Kolobov}
\affiliation{B. Verkin Institute for Low Temperature Physics and
Engineering, National Academy of Sciences, pr. Lenina 47, Kharkov
61103, Ukraine}

\author{E. Yu. Beliayev}
\affiliation{B. Verkin
Institute for Low Temperature Physics and Engineering, National
Academy of Sciences, pr. Lenina 47, Kharkov 61103, Ukraine}

\author{A. S. Panfilov}
\affiliation{B. Verkin Institute for Low Temperature Physics and
Engineering, National Academy of Sciences, pr. Lenina 47, Kharkov
61103, Ukraine}

\author{N. N. Galtsov}
\affiliation{B. Verkin Institute for Low Temperature Physics and
Engineering, National Academy of Sciences, pr. Lenina 47, Kharkov
61103, Ukraine}

\author{A. I. Prokhvatilov}
\affiliation{B. Verkin Institute for Low Temperature Physics and
Engineering, National Academy of Sciences, pr. Lenina 47, Kharkov
61103, Ukraine}

\author{J. Fink-Finowicki}
\affiliation{Institute of Physics, Polish Academy of Sciences,
32/46 Al. Lotnikov, 02-668 Warsaw, Poland}



\begin{abstract}
The ultrasonic properties of La$_{1-x}$Ca$_{x}$MnO$_{3}$
($x\approx 0.25$) with the Curie temperature $T_C$ about 200 K are
studied. Temperature dependences of longitudinal and transverse
sound velocities were measured in zero magnetic field and for
different constant magnetic fields as well. The ultrasonic study
is supported by magnetic, resistive, magnetoresistive, structural
and other measurements of the sample that facilitate
interpretation of the results obtained. The magnetic field
influence on sound properties found in this study presents some
new features of the interplay between the elastic and magnetic
properties of these compounds. It is shown that the
paramagnetic-ferromagnetic transition in the sample studied is
first order, but can become second order under the influence of
applied magnetic field.
\end{abstract}


\maketitle

\section{Introduction}
\label{int}
 The magnetic and magnetoresistive properties of mixed-valence
manganites of the type R$_{1-x}$A$_x$MnO$_3$ (where R is a
rare-earth element or La, A a divalent alkaline-earth element)
attracted much attention of scientific community in the last
decade (see reviews \cite{ramirez,coey,kim}). A huge negative
magnetoresistance (MR) near the Curie temperature of the
paramagnetic-ferromagnetic transition, $T_C$, was observed for
lanthanum manganites (R = La) with $0.2 \leq x \leq 0.5$. This
phenomenon, called ``colossal'' magnetoresistance (CMR), is viewed
as promising to advanced technology. This and other unique
properties of mixed-valence manganites are determined by the
complex spin, charge and orbital ordered phases, and, therefore,
are of a great fundamental interest for physics of strongly
correlated electrons.
\par
In spite of enormous theoretical and experimental activity in the
area of CMR manganites many questions remain unresolved. For
example, a lot of theoretical models are developed for CMR, but
they are so diverse that it is difficult to choose between them.
Moreover, a quantitative comparison of the known models with
experiment is practically impossible (or too ambiguous). For this
reason no definite consensus is reached yet about the essential
nature of CMR \cite{boris1}.
\par
It is clear that a variety of experimental methods must be used to
make progress in physics of these complex compounds. Among these
the ultrasonic investigations are very helpful ones, and were
used, therefore, fairly often, since they can give an important
information about manganites' elastic properties and how these
properties vary at magnetic transitions. Really, it is well known
that any magnetic transition or even changes in magnetic
properties with temperature or magnetic field are accompanied by
transformation or, at least, some spontaneous deformation of
crystal lattice \cite{belov}. For this reason the sound velocity
and elastic properties are rather sensitive to changes in magnetic
state. In this paper we present an ultrasonic investigation of a
bulk manganite La$_{1-x}$Ca$_x$MnO$_3$ ($x\approx 0.25$) prepared
by the floating-zone method. The study has been done at various
magnetic fields that makes the data obtained more informative.
The known ultrasonic studies of La$_{1-x}$Ca$_x$MnO$_3$ system
were usually carried out in zero magnetic field, although some
studies in magnetic fields for manganites of other types are
known (and will be discussed below).  Since any experimental
method alone is not enough for proper study of such complex
compounds, the ultrasonic study was supported by structural,
magnetic and transport measurements. All this has enabled us to
arrive at some definite conclusions about peculiarities of
magnetic states and paramagnetic-ferromagnetic (PM-FM) transition
in this type of manganites. In particular, we have shown that the
PM--FM transition is first order, but can become second order
under the influence of applied magnetic field.

\section{Experimental}
The crystal  under study was grown (at the Institute of Physics,
Warsaw) by the floating-zone method using a ceramic feed rod with
a nominal composition La$_{0.67}$Ca$_{0.33}$MnO$_3$. The crystal
growing procedure is outlined in Ref. \cite{boris2} where it was
shown that crystal perfection of samples produced by the described
technique is close to that of single crystals, and, in this
respect, they have far better crystal quality and far less
porosity than the samples prepared by a solid-state reaction
technique. On the other hand, the crystals have some extrinsic
inhomogeneties arising due to technological factors in the sample
preparation. These are rare grain boundaries and twins
\cite{boris2}.
\par
 The sample characterization and measurements were done
by a variety of experimental techniques. The structure state of
the sample was studied by the X-ray diffraction (XRD) method. The
XRD spectra (from prepared powders) were obtained using a DRON-3M
diffractometer with the $K_{\alpha}$ radiation of the Cu anode.
The errors in the intensity of diffraction peaks and the deduced
lattice parameters were 1\% and 0.02\%, respectively. The X-ray
study in a Laue camera has been done as well. The dc
magnetization was measured in a home-made Faraday-type
magnetometer. Resistance as a function of magnetic field (up to
1.7 T) and temperature (in the range 4--400 K) was measured using
a standard four-point probe technique in a home-made cryostat.
\par
Acoustic properties (sound wave velocity and attenuation) were
measured with a new version of the phase method described in
Ref.~\cite{fil}. The use of the electron-controllable phase
shifter together with the digital phasemeter permitted to gain the
resolution about one degree at practically unlimited dynamic
range. The method permits to measure the absolute values of the
sound velocities in samples on the one hand, and to register the
relative changes in the sound velocities and attenuation as a
function of temperature or magnetic field, on the other hand. The
method can give an acceptable accuracy ($\approx 1$~\%) for the
absolute value of the sound velocity in samples of a millimeter
(submillimeter) size. The standard accuracies of the measurements
for the relative changes of the sound velocity and the sound
attenuation were $10^{-4}$ and 0.05 dB, respectively.
\par
The measurements (in zero magnetic field and for different
magnitudes of the field up to 4 T) were carried out with
longitudinal and transversal waves (in the frequency range 53--55
MHz) applied to the sample studied. The magnetic field has always
been applied in the direction of the propagation vector of sound
waves.
\par
We used the sample with the dimensions $2.94\times 1.9\times 1.8$~
mm$^3$ in our acoustic investigation. The working surfaces of the
sample were carefully processed using fine abrasive powder and
were made flatly parallel (with the accuracy 1--2 $\mu$m). To
obtain the acoustic contact between piezoelectric transducers and
working surfaces of the samples we used bonding layers of the oil
GKZh-94 and the glue BF-2 for the measurements of the longitudinal
and transversal waves, respectively. The GKZh-94 is an organic
silicone polymer oil. It stays in a liquid phase till 120 K, and,
hence, an additional stress on the sample in the region of the
phase transition (200 K) is excluded. The glue BF-2 (alcohol
solution of Bakelite and polyvinylbutyral) is polymerized at the
room temperature. Since the features in the temperature behavior
of the transversal sound appear at the same temperatures, as for
the longitudinal one, we can conclude, that the bonding layers do
not significantly affect the results of the measurements.

\section{Results and discussion}
\subsection{General characterization of the sample}
The disc-shaped samples cut from the same cylindrical crystal
(about 8 mm in diameter) were used for different kinds of
investigations. The resistive and magnetoresistive properties for
one of such discs are described in Ref. \onlinecite{boris2}. It
was shown in that study, among other things, that the inner part
of the disc appeared to be more homogeneous and perfect than the
outer part. For this reason only the central part (with the
diameter about 4.5 mm) of the disc chosen for this study has been
used for ultrasonic and all other measurements described in this
paper. It turned out that the sample studied had somewhat
different magnetic and transport properties as compared with the
sample described in Ref. \onlinecite{boris2}, although both
samples were cut from the same grown cylindrical crystal. For
example, Curie temperature $T_C$ which was taken as the
temperature of the inflection point in the temperature dependence
of magnetization was equal to about 200 K in the sample studied
(Fig. 1) compared with $T_C \approx 216$ K found in Ref.
\onlinecite{boris2}. The temperature dependences of resistance
$R(T)$ and magnetoresistance (MR) are also found to be somewhat
different from that in Ref. \onlinecite{boris2} (see below).
\par
The reason for distinctions in magnetic and other properties of
the samples taken from different parts of the same grown crystal
can be considered as rather clear. It is known \cite{muk2} that
La$_{1-x}$Ca$_x$MnO$_3$ crystals grown by floating-zone technique
have an inhomogeneous distribution of La and Ca along the growth
direction. The initial part of the crystal is usually somewhat
enriched in La and depleted in Ca. The microprobe elemental
analysis of the sample in Ref. \onlinecite{boris2} had shown that
its chemical composition was close to the nominal one ($x=0.33$).
In contrast to this, the electron microprobe analysis in a
scanning electron microscope ISM-820 has shown that for the sample
studied in this work the elemental ratio Ca/(La+Ca) is about 0.25.
The ratio (La+Ca)/Mn has been found to be somewhat larger than
unity (about~1.1), that can be an evidence of some Mn deficiency.
\par
The sample was characterized also by XRD from powder. It is known
that bulk La$_{1-x}$Ca$_{x}$MnO$_3$ has a distorted perovskite
structure, which presently believed to be orthorhombic
\cite{coey,kim}. In the orthorhombic space group {\it Pnma},
lattice constants are $a \approx \sqrt{2}\,a_{p}$, $b \approx
2a_{p}$, and $c\approx \sqrt{2}\,a_p$, where $a_p$ is the lattice
constant of pseudo-cubic perovskite lattice. Generally, however,
the deviations from cubic symmetry are found to be fairly small,
especially in the CMR range ($0.2 < x < 0.5$) \cite{laiho,aken}.
It is found in this study that the powder XRD spectra of the
sample at $T=293$~K is consistent with the orthorhombic {\it Pnma}
lattice with lattice constants $a = 0.54855$~nm, $b=0.77652$~nm,
and $c=0.54994$~nm. It is seen that the orthorhombic distortions
are rather small that permits to determine from these data the
average cubic lattice constant $a_p = 0.38834$~nm. This is a
little larger than that ($a_p=0.38713$~nm) obtained in
Ref.~\onlinecite{boris2} for the sample with the composition close
to the nominal one ($x=0.33$). But this is quite expected since
$a_p$ becomes larger with decreasing Ca content $x$ in
La$_{1-x}$Ca$_{x}$MnO$_3$ manganites \cite{laiho}.
\par
The crystal perfection of the sample studied has also been proved
by X-ray study with a Laue camera. This revealed that the XRD
pattern of a small piece taken from the central part of the sample
corresponds to a single-crystal structure. The sample as a whole
is not, however, a single crystal but consists of a few grains.
\par
It is known that $T_C$ value in La$_{1-x}$Ca$_{x}$MnO$_3$ system
decreases rather sharply with decreasing $x$ in the range $0.2\leq
x \leq 0.3$ \cite{ramirez,coey,kim,rival}. In crystals prepared by
the floating-zone method the $T_C$ values equal to 216 K, 200 K
and 189 K were found for $x=0.3$ \cite{lyanda}, 0.25 \cite{aken},
and 0.22 \cite{mark}, respectively \cite{footnote}. Thus, the
value of $T_C\approx 200$~K in the sample studied is equal to that
expected for $x=0.25$ in La$_{1-x}$Ca$_{x}$MnO$_3$ crystals
prepared by floating-zone method. The same value of $T_C$ also
follows from the resistive and magnetoresistive properties of the
sample (described in the next section) which in addition to that
give evidence of rather high crystal perfection of the sample.

\begin{figure}[b]
\includegraphics[width=0.88\linewidth]{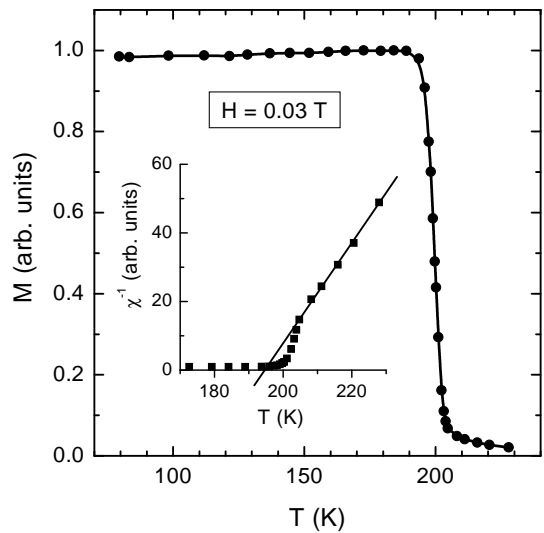}
\caption{Temperature dependence of magnetization for the
La$_{0.75}$Ca$_{0.25}$MnO$_3$ crystal (recorded in a dc magnetic
field $H=0.03$~T with increasing temperature after the sample was
cooled in zero field). The inset shows the temperature behavior of
the inverse susceptibility $1/\chi$ in the paramagnetic region at
the same field.}
\end{figure}

\par
The magnetization data (Fig. 1) allow to make some general
suggestions about the nature of the PM-FM transition in the sample
studied. It can be seen that the magnetic transition is notably
sharp and the $M(T)$ behavior obviously does not follow the
classical molecular-field theory for the PM-FM transition
\cite{belov}. The transition does not appear as a second-order
one, but more likely as a slightly broadened first-order
transition. Inset in Fig.~1 shows the temperature behavior of the
inverse susceptibility ($1/\chi=H/M$) above $T_C$. It is seen that
$1/\chi(T)$ does not follow the Curie-Weiss law in the
paramagnetic region and has a characteristic knee. This particular
kind of deviations from Curie-Weiss law (with $T_{pm} < T_C$,
where $T_{pm}$ is determined by extrapolation of linear part of
the $1/\chi(T)$ curve down to intersection with $T$-axis; $T_C$ is
the Curie temperature defined from $M(T)$ curve) had previously
been found in La$_{1-x}$Ca$_{x}$MnO$_3$ manganites with
$x\approx0.25$ or close to this \cite{zhang,amaral,salamon}, and
had been interpreted as a feature of first-order transition
\cite{amaral}. Generally, however, the nature of the ferromagnetic
transition in La-Ca manganites in the region $0.2\leq x \leq 0.3$
is still not so clear and may be influenced by intrinsic and
extrinsic inhomogeneities \cite{boris1,rival,salamon}. For
example, the above-mentioned knee-type feature in the $1/\chi(T)$
dependence above $T_C$ (like that in inset of Fig. 1) has been
attributed in Ref.~\onlinecite{salamon} to appearance of
spin-aligned (ferromagnetic) clusters above $T_C$. The cluster
size grows as the temperature becomes closer to $T_C$ (or with
increasing magnetic field) so that the magnetic transition is
governed by percolation processes (see also discussion in Ref.
\cite{boris1}).

\subsection{Resistive and magnetoresistive properties}
\label{resist}

The temperature dependences of resistivity $\rho$ and MR are shown
in Figs. 2, 3 and 4. The $\rho(T)$ curve in Fig. 2 for zero
magnetic field shows an impressively huge (50-fold) drop of the
resistivity when crossing $T_C$ from above. Note, that after this
drop the resistance continues to decrease changing further nearly
by the order of magnitude with decreasing temperature (Fig. 3). In
La-Ca manganites, $\rho(T)$ has a semiconducting behavior above
$T_C$ and metallic behavior below $T_C$ which results in a
resistivity peak at $T=T_p$ (Figs. 2 and 3). For the sample
studied, resistivity above $T_C$ (in the paramagnetic state)
follows $\rho(T)\propto \exp(E_a/T)$ with $E_a=0.1$~eV. This is in
good agreement with that (about 0.1 eV) reported by other authors
for La$_{1-x}$Ca$_{x}$MnO$_3$ crystals with nearly the same
composition \cite{coey,boris2,lyanda}. In low-temperature range,
$\rho(T)$ has a minimum at $T_{min}\approx 7.4$~K (Fig. 2).
\par
The sample possesses huge MR (Figs. 2 and 4). Taking a quantity
$\Delta R(H)/R(0) = [R(H)-R(0)]/R(0)$ as a measure of MR, the
maximum absolute values of $\Delta R(H)/R(0)$ equal to 86~\% and
91.7~\% can be determined near $T\approx T_C$ for fields 1 T and
1.5 T, respectively. This really colossal MR in not that strong
fields (note that $|\Delta R(H)|/R(0)$ by definition cannot be
higher than 100 \%) is an evidence of high crystal perfection of
the sample studied. Other important measure of crystal perfection
in manganites is the ratio, $\rho(T_p)/\rho(0)$, of peak
resistivity $\rho(T_p)$ and the residual resistivity at low
temperature ($T\approx 4$~K), $\rho(0)$. The value of
$\rho(T_p)/\rho(0)$ is about 480 for the sample studied with
$\rho(0)\approx 10^{-3}$ $\Omega$~cm.

\begin{figure}[t]
\includegraphics[width=0.89\linewidth]{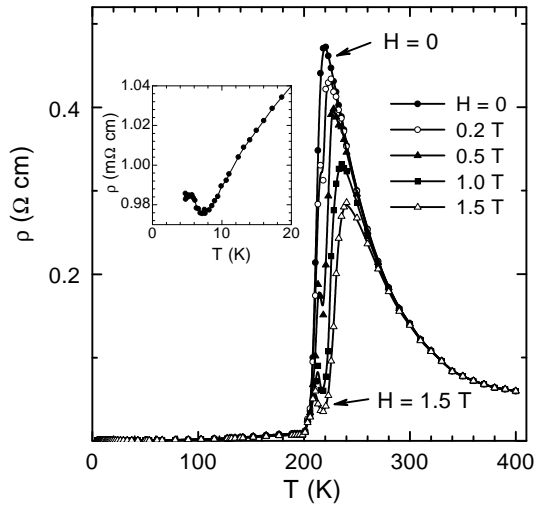}
\caption{Temperature dependences of the resistivity for the
La$_{0.75}$Ca$_{0.25}$MnO$_3$ crystal, recorded in zero magnetic
field and in fields $H=$~ 0.2, 0.5, 1, and 1.5 T. Inset shows a
shallow resistance minimum at temperature $T_{min}\approx 7.4$~K.}
\end{figure}

\par
On the whole, resistive and magnetoresistive behavior of the
sample studied (with $x\approx 0.25$) is fairly close to that of
the sample taken from another part of the grown crystal with
$x\approx 0.33$ (described in Ref. \onlinecite{boris2}). This
includes specific features of the $\rho(T)$ and $\Delta R(H)/R(0)$
dependences, such as shouldered form of $\rho(T)$ and the
low-temperature resistance minimum (Figs. 2 and 3), and
double-peaked appearance of the temperature dependences of MR for
high enough fields (Fig. 4). These features were fully considered
and elucidated in Ref. \onlinecite{boris2}, and will not be retold
in this article \cite{foot1}. It should only be mentioned that the
sharp peaks in $\rho(T)$ and MR near $T_C$ (Figs. 2 and 4) are
determined by transition from insulating paramagnetic to metallic
ferromagnetic phase in a considerable  part of the sample volume;
whereas, the shoulder in $\rho(T)$ (Fig. 3) and the second weak
peak in the temperature dependence of MR (Fig.~4) are determined
by the presence of grain-boundary-like inhomogeneities which in
the crystals grown by the floating-zone method are most likely
twin boundaries \cite{boris2,yuzhel}.
\par
The Figure 4 shows a pronounced positive MR which appears only for
low field range and peaks near $T_C$. It should be noted that the
temperature dependences of the MR in Fig.~4 were recorded for
magnetic field orientation perpendicular to the measuring current
$J$.  When applied field is parallel to the current, the MR is
always negative (compare $\Delta R(H)/R(0)$ dependences near $T_C$
for $H\| J$ and $H\bot J$ in Fig. 5). Thus, a considerable MR
anisotropy is found in the sample studied near $T_C$. This effect
has not been seen in study \cite{boris2} of other piece of the
same grown crystal because the low-field MR was not investigated
there.

\begin{figure}[t]
\includegraphics[width=0.89\linewidth]{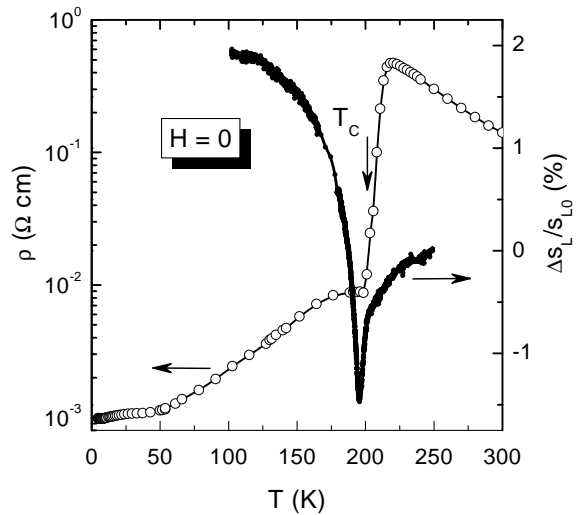}
\caption{Comparison of temperature variations of the resistivity
and the longitudinal sound velocity (taken on heating) in the
crystal studied in zero field. The position of $T_c$ is shown by
an arrow.}
\end{figure}

\par
The MR anisotropy effect found is very interesting by itself, but
is not the main goal of the present study and will be touched here
only in general. In bulk ferromagnetic samples, the dependence of
resistance on the angle between electric current and magnetic
field should actually reflect a corresponding dependence of
resistance on the angle between current and magnetization. The MR
anisotropy in the sample studied should be primarily associated
with the so called anisotropic MR (AMR) effect, determined by
spin-orbit coupling \cite{mac}. AMR is the intrinsic property of
ferromagnets and depends on the relative orientation of the
magnetization and current. The effect is well known for CMR
manganite films (see Refs. \cite{boris3,ziese,ziese2} and
references therein), but has not been mentioned previously in
known literature for bulk manganite samples.

\begin{figure}[htb]
\includegraphics[width=0.88\linewidth]{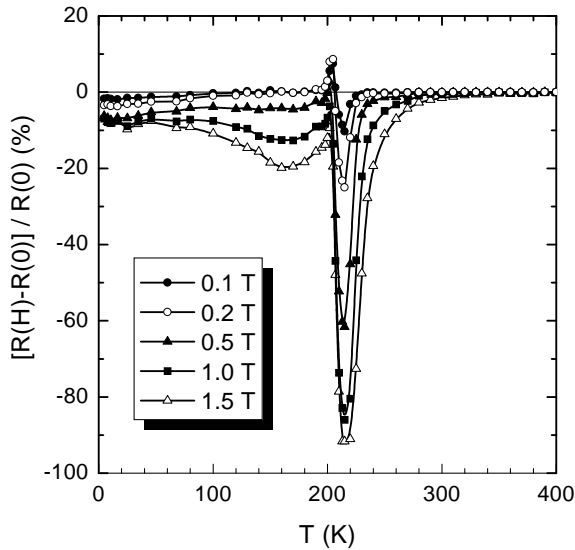}
\caption{Temperature dependences of magnetoresistance for the
sample studied, recorded at different magnitudes of applied
magnetic field.}
\end{figure}

\begin{figure}[htb]
\includegraphics[width=0.88\linewidth]{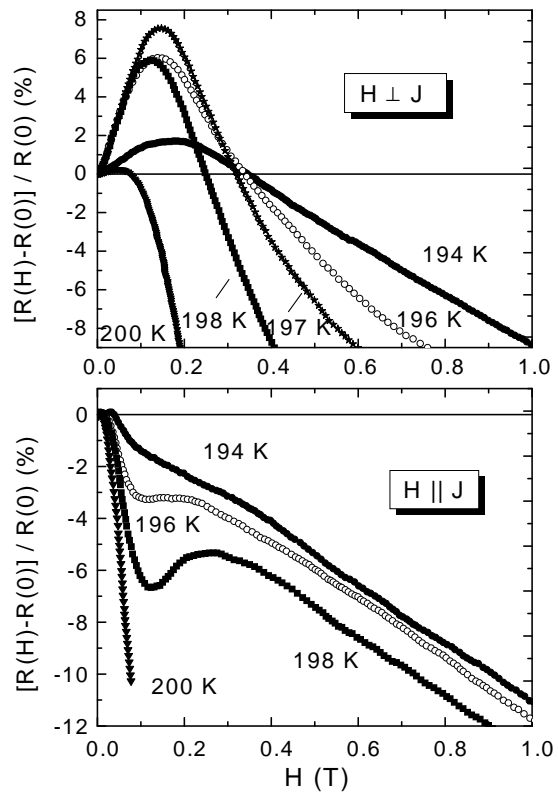}
\caption{Magnetoresistance of the sample studied for different
temperatures near $T_C\approx 200$~K for parallel and
perpendicular orientations of the magnetic field relative to the
measuring current $J$.}
\end{figure}
\par
The following quantity can be used as a measure of the AMR:
\begin{eqnarray} \delta_{an}(T,H) =
\frac{R_{\|}(T,H)-R_{\bot}(T,H)}{R_{0}(T)}, \nonumber
\end{eqnarray}
where $R_{\|}$ and $R_{\bot}$ are  longitudinal ($M\| J$) and
transverse ($M\bot J$) resistances, respectively; $R_{0}(T)$ is
zero field resistance. The applied magnetic field must be high
enough to rotate the magnetization in a selected direction. It was
found \cite{boris3,ziese,ziese2} that $\delta_{an}$ is negative
for manganite films (in contrast to $3d$ metals and alloys where
it is usually positive) and is a maximum at $T\approx T_C$. In
this regard, the MR anisotropy found in this study for the bulk
manganite crystal behaves like that for manganite films. There is,
however, a significant difference in the magnitude of the effect.
The maximum absolute value of $\delta_{an}$ near $T_C$ for the
films is 1--2~\% \cite{boris3,ziese,ziese2}. This rather small
magnitude (in comparison with that known for 3$d$ metals
\cite{mac}) was even justified for manganites in the frame of the
approach developed for 3$d$ metals \cite{ziese}. In the sample
studied, the maximum magnitude of $\delta_{an}$ (about 15 \%, as
can be estimated from Fig. 5) is, however, far higher. It should
be noted that in ferromagnetic films, beside the AMR effect, some
further sources of MR anisotropy are inevitably present
\cite{boris3}, for example, the existence of preferential
directions of magnetization due to strains stemming from the
lattice film-substrate mismatch, the influence of the shape
anisotropy and other sources. In bulk samples these factors are
either absent or not so important. Magnetocrystalline anisotropy
in manganites with nearly cubic-lattice symmetry \cite{laiho,aken}
is negligible and, therefore, cannot exert much influence on MR
anisotropy in bulk samples. It is known as well that the AMR
effect in manganite films becomes apparent only in samples with
fairly high crystal perfection \cite{ziese2}. Maybe for some of
these reasons $\delta_{an}$ is so high in the rather perfect bulk
crystal studied. It can be said, to summarize, that the problem of
the AMR in manganites cannot be considered as quite clear and
further theoretical and experimental work in this direction is
needed.

\subsection{Ultrasonic properties near the PM-FM transition}

Before presenting ultrasonic results obtained, we will mention
briefly the known ultrasonic studies of the PM-FM transition in
La$_{1-x}$Ca$_{x}$MnO$_3$ system. It was generally revealed (on
ceramic samples with chemical composition similar to that of the
sample studied; see, for example, Refs.
\onlinecite{ramirez,ramirez2,zhu1}) that both, the longitudinal
and transverse sound velocities begin to increase continuously
when crossing $T_C$ from above, amounting to values up to $\approx
5$ \% larger than those above $T_C$. The sound attenuation shows
a rather sharp peak at $T\approx T_C$. An external magnetic field
suppresses this sound anomaly in the neighborhood of $T_C$
\cite{ramirez,ramirez2}.  In other studies \cite{zhu2,fuji1} of
La-Ca manganites (made as we believe on samples with higher
crystal perfection) the temperature dependence of sound velocity
was found to reveal a weak but quite clear minimum near $T_C$,
that is, with decreasing temperature the sound velocity first
decreases noticeably when approaching $T_C$ (so called, softening)
and then increases after crossing $T_C$. The same behavior was
found in La-Sr manganite (La$_{0.67}$Sr$_{0.33}$MnO$_3$) near the
PM-FM transition \cite{rajen}.
\par
The sample studied, although being rather perfect, consists,
however, of large crystals with different orientations, so that
our ultrasonic study is carried out as that for a polycrystalline
sample. Temperature dependences of the longitudinal sound velocity
$s_L$ and attenuation $\alpha$ measured in this work (Figs. 3 and
6) show clear anomalies in the region near $T_C\approx 200$~K: a
minimum in $s_{L}(T)$ and a peak in the attenuation, that agrees
with data of Refs. \onlinecite{zhu2,fuji1}. The minimum in
$s_{L}(T)$ appears to be, however, much more expressive and
sharper as compared with that in Refs. \onlinecite{zhu2,fuji1}. In
this respect, this anomaly is more similar to a considerable sound
velocity softening found in the vicinity of the charge-ordering
transition temperature $T_{CO}$ in La$_{1-x}$Ca$_{x}$MnO$_3$ for
Ca concentrations around $x\approx 0.2$ and $0.5 \leq x \leq 0.82
$ \cite{ramirez,ramirez2,zhu1,fuji2}. According to Ref.
\cite{fuji2}, it should be attributed to large local volume
fluctuations which are caused  by the two-phase competition near
the transition temperature. The same reason can be responsible for
the sharp minimum in $s_L(T)$ near $T_C$ under the PM-FM
transition in the sample studied.

\begin{figure}[htb]
\includegraphics[width=0.87\linewidth]{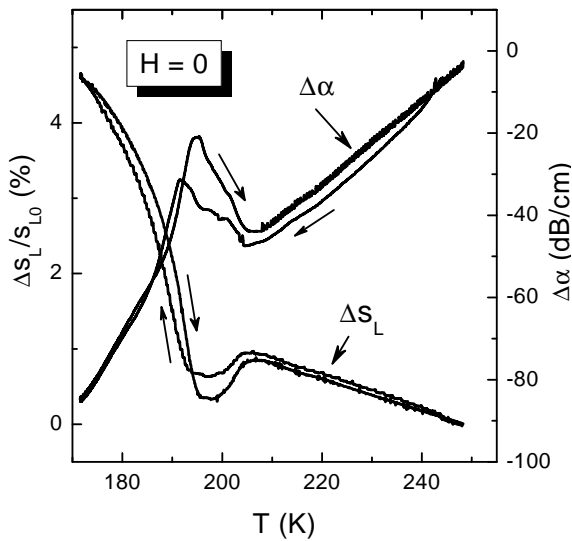}
\caption{Relative changes in the longitudinal sound velocity $s_L$
and sound attenuation $\alpha$ in the region around Curie
temperature $T_C \approx 200$~K at zero magnetic field. The arrows
near the curves show directions of temperature variations during
measurements (heating and cooling). Both, heating and cooling
rates were about 0.16 K/min.}
\end{figure}

\par
The $s_L(T)$ and $\alpha(T)$ dependences are hysteretic in a
rather wide region around the PM-FM transition. This clearly
indicates that the transition is more in the nature of first-order
one, that also follows from the temperature behavior of the
inverse susceptibility $1/\chi$ in the paramagnetic region (inset
in Fig. 1). The temperature dependence of the transverse sound
velocity $s_T$ (Fig. 7) shows an increased slope in the region of
$T_C$ with a considerable change in the slope below $T_C$, but
generally the $s_T(T)$ curve does not demonstrate some clear
features around $T_C$.
\par
The Figures 6 and 7 show the relative temperature variations of
the longitudinal $s_L$ and transverse $s_T$ sound velocities. The
absolute values of the sound velocities have been measured at
$T=77$~K. The following values were obtained: $s_L=6.4\times
10^5$~cm/s and $s_T=2.65\times 10^5$~cm/s for the sample with
acoustical path 2.94 mm. They agree fairly well with those
obtained by other researchers in La-Ca manganites of the
approximate chemical composition \cite{fuji2,mira}. Figure~3
demonstrates a comparison of temperature variations of the
longitudinal sound velocity and the resistivity (both taken on
heating) in the crystal studied. It is seen that the temperature
position of the sound anomaly agrees closely with sharp drop in
the resistance near $T\approx T_C$.

\begin{figure}[t]
\includegraphics[width=0.85\linewidth]{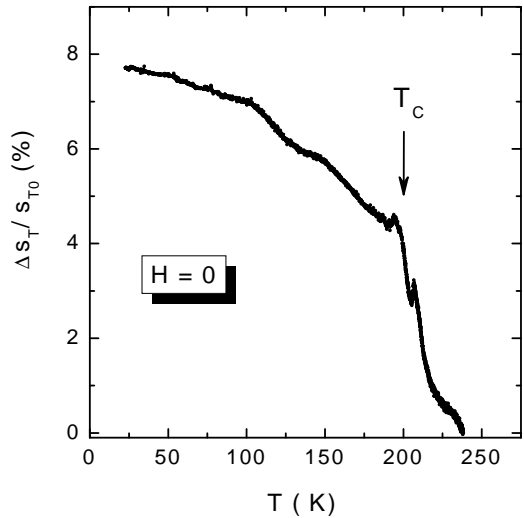}
\caption{Relative changes in the transverse sound velocity $s_T$
in the region around Curie temperature $T_C \approx 200$~K at zero
magnetic field. The dependence was taken on heating with rate
about 0.16 K/min.}
\end{figure}

\par
The influence of magnetic field on the $s_L(T)$ anomaly in the
region of PM-FM transition is shown in Fig.~8. It is seen that
the magnetic field up to 4 T does not suppress the sound-velocity
anomaly significantly, opposed to the data of Refs.
\onlinecite{ramirez,ramirez2}. As a matter of fact, in our study
the magnitude of the anomaly increases considerably and its width
becomes broader with increasing applied field up to 2 T, and only
in the highest applied field ($H=4$~T) it reduces to some degree
being still larger than that at zero field. The second important
feature is that the temperature position of the anomaly (as a
measure of which the temperature $T_{tr}$ of the minimum in
$s_L(T)$ can be taken) moves steadily to higher temperature with
increase in an external field. This is shown more clearly in Fig.
9 where magnetic-field dependence of the transition temperature
$T_{tr}$ is presented. Due to the considerable hysteresis
(Fig.~6), the values of $T_{tr}(H)$ taken from the $s_L(T)$ curves
on heating and cooling are considerably different. This difference
reduces, however, nearly to zero with increasing magnetic field up
to $H=4$~T (Fig.~9). The attenuation peak for the longitudinal
sound behaves in magnetic field in the same way as the
sound-velocity anomaly: its position shifted steadily to higher
temperature with increasing field, and the peak's magnitude
depends on the field in the same non-monotonic way.

\begin{figure}[t]
\includegraphics[width=0.88\linewidth]{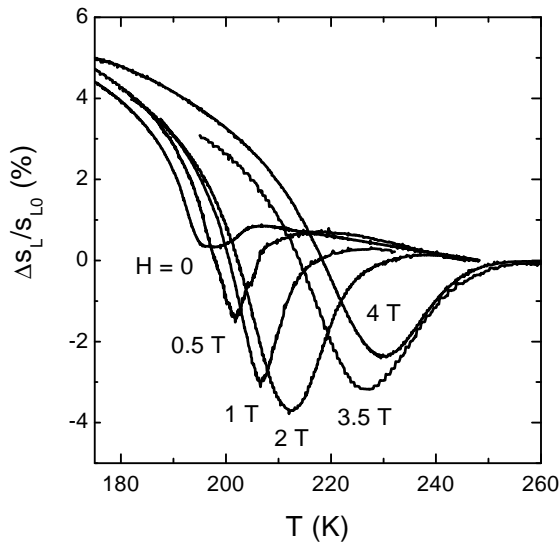}
\caption{Relative changes in longitudinal sound velocity $s_L$ in
temperature range near the paramagnetic-ferromagnetic transition
in different magnetic fields. All curves (except at $H=3.5$~T)
were taken on heating.}
\end{figure}

\par
The magnetic-field dependences of the longitudinal and transverse
sound velocities are presented in Figs. 10 and 11. For
longitudinal waves the magnetic field behavior of the relative
quantity  $[s_{L}(H)-s_{L}(0)]/s_{L}(0)$ (with $s_{L}(0)$ being
the sound velocity in zero field at room temperature) is presented
in Fig.~10. The similar quantity, $[s_{T}(H)-s_{T}(0)]/s_{T}(0)$,
is used for the transverse velocity in Fig.~11., where the curves
are somewhat shifted relative to each other to show more clearly
the behavior of this parameter for different temperatures. These
parameters present well-define magnetosonic effect (MSE), which
characterizes the influence of magnetic field on the longitudinal
and transversal waves (much as, for example, the magnetoresistive
effect is a similar characteristic for the resistivity). The
longitudinal MSE is positive below $T_C$ where it has a monotonic
dependence on magnetic field (Fig. 10). It is very small for $T\ll
T_{C}$, but increases with increasing temperature when going to
$T_C$. But the behavior of the longitudinal MSE changes
dramatically when temperature crosses $T_C$ (Fig. 10). It becomes
negative for $T\geq T_C$ and its behavior in magnetic field
becomes non-monotonic. The field position of the minimum in MSE in
this temperature range increases with increasing temperature.
Actually, the ($T,H$) coordinates of these minima correspond
rather closely to the phase diagram $T_{tr}$-$H$ shown in Fig. 9.
This indicates that the shift of the temperature $T_{tr}$ in
magnetic field and shift of the magnetic-field position of the
minimum in the MSE with increase in temperature are determined by
processes of the same origin. These are related to magnetoelastic
(or magnetostructural) interaction in the sample studied and will
be discussed below. The transverse MSE (Fig. 11) behaves quite
differently than the longitudinal one. It has a non-monotonic
(with minimum) magnetic-field behavior below $T_C$ and the
monotonic one above it. In the same manner as the longitudinal
effect, the transverse MSE is negligible at low temperature $T\ll
T_C$, but increases considerably with increasing temperature when
moving to $T_C$.

\begin{figure}[h]
\includegraphics[width=0.85\linewidth]{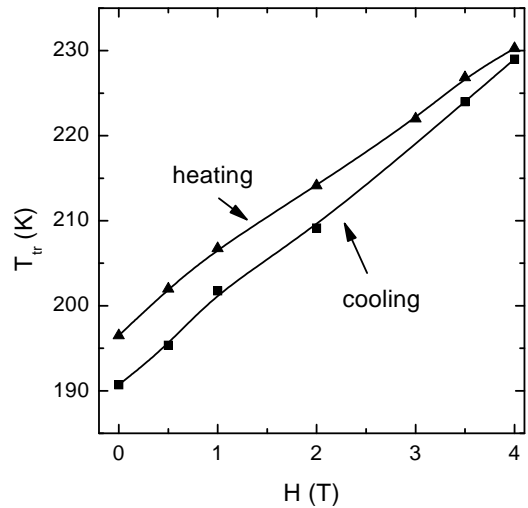}
\caption{Influence of magnetic field on the transition temperature
$T_{tr}$.  The temperature $T_{tr}$ is defined as a position of
the temperature minimum in $s_L(T)$ curves (Fig. 8) in the region
of the PM-FM transition. The values of $T_{tr}(H)$ were taken from
$s_L(T)$ curves measured in different magnetic fields on heating
and cooling.}
\end{figure}

\par
The presently attained knowledge of the CMR manganites is quite
enough for (at least) qualitative interpretation of the results
obtained. The available literature indicates that significant
changes in the crystal-lattice and elastic properties of La-Ca
manganites take place under the PM-FM transition or with
strengthening of the FM order by an external influence. For
example, the PM-FM transition in ceramic samples of
La$_{1-x}$Ca$_{x}$MnO$_3$ ($0.25\leq x\leq 0.35$) is known to be
accompanied by rather significant contraction of the crystal
lattice, which causes a peak in the thermal expansion coefficient
near $T_C$ \cite{radael,dai,ibarra,huang,hibble}. Above $T_C$, the
lattice expansion is enhanced compared with prediction of the
Gr\"{u}neisen law. The large changes in the lattice unit-cell
volume at the PM-FM transition is a doubtless evidence that this
is a magnetostructural transition of first order.  It is found for
a La-Ca manganite sample with $x=0.33$ \cite{huang} that the
increase in magnetic field in the temperature range near $T_C$
(including $T$ somewhat below or above it) causes a decrease in
the unit-cell volume together with simultaneous increase in the
saturated magnetization (in the region of the so called
paraprocess \cite{belov}). This shows clearly that strengthening
of magnetic order in manganites results in contraction of the
crystal lattice (and vice versa, as the studies of the pressure
effect on $T_C$ show \cite{coey}).
\par
The magnetostriction  study of Ref. \onlinecite{ibarra} for La-Ca
manganite with $x=0.33$ appears to be helpful for understanding of
ultrasonic data of this study. The authors of that study have
measured longitudinal $(\Delta l/l)_\|$, transverse $(\Delta
l/l)_\bot$ and volume $\Delta V/V=(\Delta l/l)_\| +2(\Delta
l/l)_\bot$ magnetostrictions (MS), and found that $\Delta V/V$ is
negative in a wide range around $T_C$ with a maximum of its
absolute value peaked at $T\approx T_C$. But for $T\leq T_{C}/2$
it becomes positive. The MS is isotropic [$(\Delta l/l)_\|\approx
\Delta l/l)_\bot$] in a rather wide range above $T_C$ and in some
narrower range below it. When moving further below $T_C$, the
temperature behaviors of the longitudinal and transverse MS
become, however, quite different: $(\Delta l/l)_\|$ changes sign
and becomes positive of an appreciable value with decreasing
temperature; whereas, $(\Delta l/l)_\bot$, being negative,
approaches nearly zero value with decreasing temperature, and only
at low temperature (below 50 K) it attains a rather small positive
value.

\begin{figure}[h]
\includegraphics[width=0.88\linewidth]{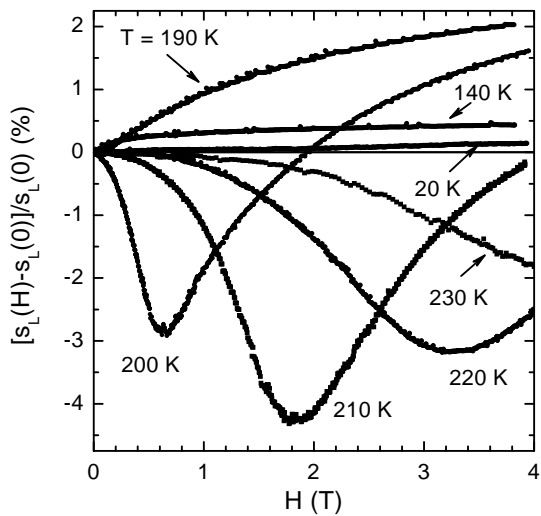}
\caption{Relative dependences of the longitudinal sound velocity
on applied magnetic field recorded for different constant
temperatures.}
\end{figure}

\par
Beside the appreciable lattice volume  changes, noticeable changes
in local lattice structure take place as a result of the PM-FM
transition or with strengthening of the FM order under the
influence of magnetic field
\cite{aken,dai,huang,hibble,radael2,billi,li}. Bulk
La$_{1-x}$Ca$_x$MnO$_3$ has a distorted perovskite structure,
which (to a fairly good approximation) is orthorhombic. The
orthorhombic deformation of the cubic perovskite lattice is
determined by, the so called, GdFeO$_3$ rotation of MnO$_6$
octahedra and the Jahn-–Teller (JT) distortion \cite{kim,aken}.
For the concentration range $0.25\leq x\leq 0.5$, the JT
distortion is found to be rather small \cite{hibble,aken}, but
cannot be entirely neglected. In mixed-valence manganites,
containing JT Mn$^{3+}$ ions, the local JT distortions of MnO$_6$
octahedra should be taken into account, even if the average JT
distortion is negligible. The existing data on this matter are
still rather incomplete and discrepant, but they allow
nevertheless to suppose that transition from the PM to FM state
results in decrease in the orthorhombic and JT distortions. For
example, in the La-Ca manganites with $0.25\leq x \leq 0.33$ the
JT distortion decreases considerably when system going from the PM
to FM state \cite{hibble,radael2,billi}. The Mn-O-Mn bond angle
increases \cite{radael2}; whereas, GdFeO$_3$ rotation of MnO$_6$
octahedra decreases \cite{aken} in the FM state, that testifies to
lessening of the orthorhombic distortions in this state. Since
undistorted MnO$_6$ octahedra are ascribed to the metallic FM
phase; whereas, the JT distorted octahedra are attributed to the
PM insulating phase, it is reasonably suggested that in the
transition region on both sides of $T_C$ the charge localized and
delocalized phases coexist \cite{billi}. The strengthening of the
magnetic order above $T_C$ induced by external field about 2 T
causes decrease in the orthorhombic and JT distortions in
La$_{0.75}$Ca$_{0.25}$MnO$_3$ \cite{li}. It can be concluded,
therefore, that enhancing of the magnetic order for any reason
leads to relief of the orthorhombic and JT distortions.

\begin{figure}[h]
\includegraphics[width=0.85\linewidth]{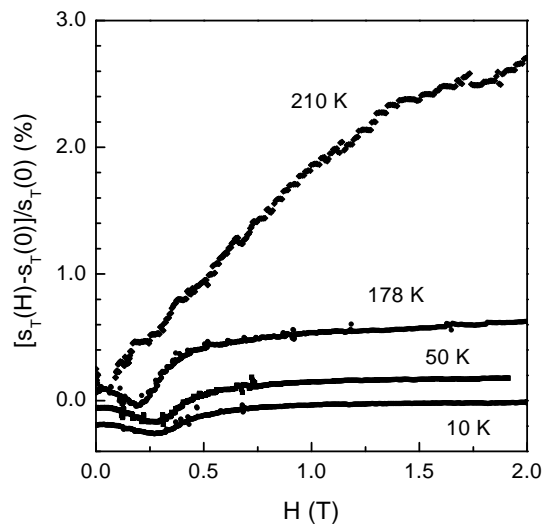}
\caption{Relative dependences of the transverse sound velocity on
applied magnetic field recorded for different constant
temperatures. }
\end{figure}

\par
It is known that fluctuations of the magnetic order and lattice
volume near magnetic transitions cause increased absorption and
hindered propagation of the sound at the Curie and N\'{e}el
temperatures. For this reason, in many magnetic materials  a sharp
peak in the ultrasonic attenuation together with a sharp dip in
the longitudinal sound velocity are frequently observed near the
temperature of magnetic transitions \cite{bennet,luthi,tachiki}.
For the transitions, which can be considered as second order, the
inhomogeneous magnetic state near the critical point $T_C$ is
thought to be determined by the thermodynamic fluctuations of the
spin order \cite{belov}. The PM-FM transition in the sample
studied is undoubtedly first order. An inhomogeneous magnetic
state near $T_C$ is inherent for this type of phase transition as
well. In La-Ca manganites with $0.25\leq x \leq 0.5$, the
multiphase coexistence near $T_C$ can be determined by various
reasons \cite{coey,boris1,billi,papa}. Beside the PM and FM
phases, this multiphase state can include even nanoclusters of the
antiferromagnetic phase \cite{garcia}. It is found, in particular,
that FM clusters are present well above $T_C$ while some PM
insulating clusters can persist down to a range far below $T_C$.
This implies that the PM-–FM transition has a percolative
character. With decreasing temperature, the PM volume fraction
decreases and that of the FM fraction increases. Since the PM
phase is insulating and the FM one is metallic, some kind of
insulator-–metal transition takes place near $T_C$. Beside this
the PM phase has an increased lattice unit-cell volume. All these
factors cause anomalies in ultrasonic properties near $T_C$.
\par
The ultrasonic anomalies near $T_C$ in the
La$_{0.75}$Ca$_{0.25}$MnO$_3$ manganite found in this study are
determined by magnetoelastic (spin-phonon) interaction. The
existent theories of these anomalies are developed primarily for
second order transitions \cite{bennet,luthi,tachiki}, but their
main ideas can be used for first order transitions as well.
Besides, the latent heat for the PM-FM transition in La-Ca
manganites of the approximate composition is rather small
\cite{kim3,adams} so that this transition is not so far from
second order one.  According to the models
\cite{bennet,luthi,tachiki}, the spin-phonon interaction
responsible for the ultrasonic anomalies in most cases comes from
the strain modulation of the exchange interaction (volume
magnetostrictive coupling). Although these models take into
account the linear MS as well, we will consider in good
approximation only the influence of the volume MS which dominates
over the linear one for any ferromagnet  near $T_C$ \cite{belov}.
This is true for La-Ca manganites as well \cite{ibarra}. In this
case, according to Refs. \onlinecite{bennet,luthi}, the
longitudinal sound velocity should show a dip near $T_C$
independent of the direction of propagation for both the isotropic
and anisotropic materials. A situation with the transversal waves
is more delicate. It is predicted in general \cite{bennet,luthi}
that the transverse ultrasonic anomaly should be less in magnitude
than that for the longitudinal waves. Besides, the transverse-wave
effect is significantly anisotropic being very small for some
directions. In polycrystalline samples, therefore, no large
transverse effect can be expected. Maybe for this reason the
temperature behavior of the transverse sound does not reveal some
distinct anomaly near $T_C$ in contrast to the longitudinal waves
(Figs.~6 and 7).
\par
For an isotropic solid-state medium (as the polycrystalline sample
studied with the pseudocubic perovskite lattice can be considered
in the first approximation) the following expressions for the
longitudinal and transverse sound velocities are known:
\begin{eqnarray} s_{L}(T,H) =
\sqrt{\frac{E(1-\sigma)}{d(1+\sigma)(1-2\sigma)}}=\sqrt{\frac{K+4G/3}{d}},
\end{eqnarray}
and
\begin{eqnarray} s_{T}(T,H) =\sqrt{\frac{E}{2d(1+\sigma)}}=
\sqrt{\frac{G}{d}},
\end{eqnarray}
where $E$ is the Young's modulus, $\sigma$ the Poisson's ratio,
$K$ the bulk modulus, $G$ the shear modulus, $d$ the mass density.
The magnitudes of the moduli $E$, $G$, $K$ and that of the ratio
$\sigma$ are connected by the relations
\begin{eqnarray} G = \frac{E}{2(1+\sigma)}, \qquad K = \frac{E}{3(1-2\sigma)},
\end{eqnarray}
so that, actually, the velocity $s_L$ is determined by two
independent moduli only; whereas, $s_T$ depends on the shear
modulus G alone. The reciprocal of the bulk modulus $K$ is the
compressibility. It is clear that great softening of the
longitudinal velocity near $T_C$ found in this study (Figs. 3 and
8) is connected with decrease in the moduli's magnitude (or
increase in the compressibility, which usually peaks at the PM--FM
transition).
\par
Consider more closely the effect of magnetic field on the
longitudinal sound propagation. It is seen that the $s_{L}(T)$
anomaly widens in magnetic field and its temperature position
($T_{tr}$) moves to higher temperature with increasing field
(Figs. 8 and 9). With decreasing temperature from above $T_C$ the
longitudinal velocity $s_L$ softens due to developing the
inhomogeneous magnetic state when moving closely enough to $T_C$.
After crossing $T_C$, the velocity $s_L$ start to increase as the
system becomes more homogeneous after transition to the FM state
and arrives at far more magnitude than that above $T_C$ (Fig. 3).
It is evident that the temperature $T_{tr}$ indicates, actually,
the point of the most inhomogeneous state, and this temperature
can be taken as the temperature of the PM--FM transition in the
zero and non-zero magnetic fields. It is quite expected that the
temperature of the first-order PM--FM transition is shifted under
the influence of magnetic field. This is well known for La-Ca
manganites as well. For example, $T_C$ of
La$_{1-x}$Ca$_{x}$MnO$_3$ ($x=0.33$) with $T_C\approx 260$~K in
zero field moves to higher values in magnetic fields with the rate
$dT_{C}/dH = 19$ K/T \cite{kim3}. For the sample studied this
quantity (taken on cooling) is about 10 K/T (as can be easily
estimated from Fig.~9). This is somewhat less than that obtained
in Ref. \onlinecite{kim3}, but for the sample with smaller values
of Ca concentration ($x\approx 0.25$) and $T_C$ ($\approx 200$~K)
this fairly comes up to expectations. It should be pointed out
that the temperature position of the sharp resistance drop,
corresponding to the PM--FM transition, moves to higher
temperature as well with increase in magnetic field (Fig. 2), so
that, in this study, the resistive and magnetoresistive data agree
fairly well with the ultrasonic ones.
\par
Due to considerable hysteresis (Fig.~6), the values of $T_{tr}(H)$
taken from the $s_L(T)$ curves on heating, $T_{tr}^{up}$, and
cooling, $T_{tr}^{dn}$, are essentially different, as expected for
first order transition. The difference ($\Delta T_{tr} =
T_{tr}^{up} -T_{tr}^{dn}$) reduces, however, nearly to zero with
increasing magnetic field up to $H=4$~T (Fig.~9). It appears that
the transition changes from  first to second order under the
influence of external magnetic field. The case, when a first order
transition becomes second, as an external parameter is varied, is
considered in Ref. \onlinecite{imry}. This theory takes into
account that the coherence length $\xi$ remains finite at a first
order transition. If this length is sufficiently large near the
phase transition to ''average out`` the existing (intrinsic and
extrinsic) inhomogeneities, the transition is sharp, as should be
for a first order transition. If, on the other hand, the growth of
correlations with moving to the transitions is blocked by
inhomogeneities  of different kind, the transition becomes
''smeared out`` and looks like a second order (the same effect can
be induced by extrinsic inhomogeneities \cite{boris1}). The
application of magnetic field can increase the degree of magnetic
inhomogeneity in manganites (enhancing the mixed-phase state near
the transition point) for different reasons \cite{boris1} and
leads, therefore, to significant rounding of a transition, so that
at a sufficiently high field a first order transition appears as
(or becomes) second order. It is really seen in Fig. 8 that the
$s_L$ anomaly becomes broader and more smeared with increasing
field, that corresponds to the pattern outlined in Ref.
\cite{imry}.
\par
The magnetic behavior of $\Delta T_{tr}= T_{tr}^{up} -T_{tr}^{dn}$
found in this study corresponds in general to the main conception
of Ref. \onlinecite{chad} for disorder-broadened first order
transitions. The authors of that work suppose generally that
$\Delta T_{tr}$ (which they called the temperature window of
transition) must decrease (increase) with increasing some
influencing parameter for transitions in which $T_C$ rises (falls)
with that parameter. This statement is true for our case if the
magnetic field is taken as the influencing parameter. It is
interesting to note that this rule is also true for the manganites
(for example, Nd$_{1/2}$Sr$_{1/2}$MnO$_3$ and others
\cite{kaplan}) which undergo the transition from the FM to
antiferromagnetic (or charge-ordered) state with decreasing
temperature. In this case the application of magnetic field causes
decrease in $T_N$ (or $T_{CO}$) which is accompanied by increase
in $\Delta T_{tr}$. This effect in Nd$_{1/2}$Sr$_{1/2}$MnO$_3$ for
the charge-ordering transition has found it's confirmation in
ultrasonic study \cite{zvyagin} as well .
\par
The Figure 10 demonstrates that the longitudinal sound velocity
$s_L$ is fairly sensitive to changes in magnetic state of the
manganites induced by magnetic field. As indicated above, a
strengthening of FM order (and corresponding decrease in the
specific volume) leads to an increase in $s_L$. The ability of an
external magnetic field to increase the magnetization can explain
in general terms the behavior $s_{L}(H)$ below $T_C$ (Fig. 10). It
is clear that at low temperature ($T\ll T_C$) when nearly all
spins are already aligned by the exchange interaction, this
ability is minimal. For this reason, the longitudinal MSE is very
weak for $T\ll T_C$, but is enhanced profoundly with increasing
temperature remaining positive and monotone below $T_C$. The
things get changed when temperature rises close enough to and
above $T_C$. Here, on the one hand, the possibility to strengthen
the magnetic order with an external magnetic field increases
profoundly, since the magnetic order becomes weaker; on the other
hand, however, the application of magnetic field increases
inhomogeneity in manganites (enhancing the mixed-phase state near
transition) and leads to significant rounding of the transition.
The competition of these two mechanisms results in the minimum in
$s_{L}(H)$ in this temperature range (Fig. 10). With increasing
field, $s_{L}(H)$ at first decreases due to developing
inhomogeneous magnetic state (inducing the negative MSE) and then,
with further increasing field, it rises due to strengthening of
the FM order (increase in the saturation magnetization in the
paraprocess).
\par
The temperature behavior of the transverse velocity $s_{T}(T)$
does not show any pronounced anomaly in the region of PM--FM
transition (Fig. 7). The same is true for magnetic-field behavior
$s_{T}(H)$ near and above $T_C$ (Fig. 11). Below $T_C$ the
behavior of $s_{T}(H)$ is non-monotonic with a minimum in the
field region of the technical saturation of the magnetization.
According to our data, this saturation takes place at
$H_{s}\approx 0.03$~T below 100 K. Above $H_s$ (that is in the
paraprocess) the transverse MSE is positive (Fig. 11). The field
$H_s$ goes to zero with temperature rising to $T_C$, so that the
minimum in $s_{T}(H)$ disappeares near $T_C$.
\par
In conclusion, the peculiarities of the PM--FM transition in the
sample of La$_{1-x}$Ca$_{x}$MnO$_3$ ($x=0.25$) prepared by the
floating-zone method have been studied by transport, magnetic and
ultrasonic measurements. In particular, a sharp peak in the
ultrasonic attenuation together with a very sharp dip (softening)
in the longitudinal sound velocity $s_L$ have been found near the
Curie temperature. This is in contrast with previous ultrasonic
studies of manganites of the approximate composition where mainly
only a pronounced increase in $s_L$ \cite{ramirez,ramirez2,zhu1})
or very slight dip \cite{zhu2,fuji1} as a result of the PM--FM
transition was found. We attribute this to a rather disordered
polycrystalline state of the samples studied in those references
that resulted in ``smearing out'' the magnetic transition. Our
results indicate that the PM--FM transition in the sample studied
is first order. This transition, however, can become second order
under the influence of the high enough applied magnetic field
which increases a degree of inhomogeneity in the region near
$T_C$. This behavior corresponds in general to the known
theoretical model \cite{imry}.
\par
We wish to thank V. D. Fil and A. A. Zvyagin for helpful
suggestions and V. M. Stepanenko (Institute for Single Crystals,
National Academy of Sciences, Kharkov, Ukraine)  for compositional
analysis of the sample studied.



\begin{thebibliography}{00}
\bibitem{footnote}It follows from the known phase diagram of the
La$_{1-x}$Ca$_{x}$MnO$_3$ \cite{ramirez,coey,kim} that $T_C$
values should be about 250--260 K for $0.3\leq x\leq 0.35$ with
lesser values (220--225 K) at $x=0.25$. This diagram was obtained,
however, for ceramic samples prepared by the solid-state reaction
method. In more perfect samples prepared by the floating-zone
method, $T_C$ values for $0.22 \leq x \leq 0.35$ are usually
smaller  by 20--30 K (see discussion in Ref.~\onlinecite{boris2}).
\bibitem{foot1}We can point out for comparison some parameters for the
sample in Ref. \onlinecite{boris2}: $\rho(T_p)/\rho(0)\approx
200$, the activation energy in paramagnetic state $E_a\approx
0.09$~eV, resistance minimum is at $T_{min}\approx 16$~K, and
maximum $\Delta R(H)/R(0) $ are 84.4 \% and 93.1 \% at $H=$~1 T
and 2 T, respectively. It seems that the sample studied appears
not less (or even more) perfect than that in Ref.
\onlinecite{boris2}.
\end{thebibliography}

\end{document}